\documentstyle[12pt]{article}
\makeatletter
\def\@maketitle{\newpage
 \null
 {\normalsize \tt \begin{flushright}
  \begin{tabular}[t]{l} \@date
  \end{tabular}
 \end{flushright}}
 \begin{center}
 \vskip 2em
 {\LARGE \@title \par} \vskip 1.5em {\large \lineskip .5em
 \begin{tabular}[t]{c}\@author
 \end{tabular}\par}
 \end{center}
 \par
 \vskip 1.5em}
\makeatother
\begin{document}
\setlength{\baselineskip}{16pt}
\title{Quantum $R^2$ Gravity in Two Dimensions}
\author{
        Hikaru KAWAI${}^1$\thanks{
                          kawai@tkyvax.phys.s.u-tokyo.ac.jp}
       ~~and~~
        Ryuichi NAKAYAMA${}^2$\thanks{
                          nakayama@theory.kek.jp}
\\[1cm]
{\small
    ${}^1$\, Department of Physics, University of Tokyo,} \\
{\small    Bunkyo-ku, Tokyo 113, Japan} \\
{\small
    ${}^2$\, National Laboratory for High Energy Physics(KEK),} \\
{\small
                     Tsukuba-shi, Ibaraki 305, Japan}
}
\date{
  KEK-TH-355 \\ KEK preprint 92-212 \\ UT-634 \\
  February 1993
}

\maketitle

\begin{abstract}

Two-dimensional quantum gravity with an $R^2$ term is investigated in the
continuum framework.  It is shown that the partition function for
small area $A$ is highly suppressed by an exponential factor
$exp \{ -2\pi (1-h)^2/(m^2A) \}$, where $1/m^2$ is the coefficient (times
$32\pi$) of $R^2$ and $h$ is the genus of the surface.
Although positivity is violated, at a short distance scale ( $\ll 1/m$)
surfaces are smooth and the problem of the branched polymer is avoided.

\end{abstract}

\newpage
\newtheorem{lemma}{Lemma}
\newtheorem{thm}[lemma]{Theorem}
\newtheorem{alemma}{Lemma}
\newtheorem{athm}[alemma]{Theorem}

\newcommand{\proof}{\noindent{[{\it proof\,}]}}
\newcommand{\bb}{\vrule height8pt width3pt depth0pt}
\newcommand{\remark}{\noindent{[{\it remark\,}]}}

\newcommand {\comb}[2]{{{#1} \choose {#2}}}

\newcommand {\n}{\nonumber\\}
\newcommand {\nn} {\nonumber}

\newcommand {\cleqn}{\setcounter{equation}{0}}
\renewcommand {\theequation}{\arabic{equation}}
\newcommand \eq[1]{(\ref{#1})}

\newcommand {\beq}{\begin{equation}}
\newcommand {\eeq}{\end{equation}}

\newcommand {\beqa}{\begin{eqnarray}}
\newcommand {\eeqa} {\end{eqnarray}}

\newcommand {\beqc}{\beq\begin{array}{c}}
\newcommand {\eeqc}[1]{\label{#1}\end{array}\eeq}

\newcommand {\beql}{\beq\begin{array}{l}}
\newcommand {\eeql}[1]{\label{#1}\end{array}\eeq}

\newcommand {\beqr}{\beq\begin{array}{r}}
\newcommand {\eeqr}[1]{\label{#1}\end{array}\eeq}

\newcommand {\lfrac}[2]{\frac{\displaystyle #1}{\displaystyle #2}}
\newcommand {\sfrac}[2]{\displaystyle \frac{#1}{#2}}

\newcommand {\dee}{\partial}
\newcommand {\deeb}{\bar{\partial}}
\newcommand {\del}[1]{\partial_{#1}}
\newcommand {\deldel}[1]{\frac{\delta}{\delta {#1}}}
\newcommand {\ldeldel}[1]{\lfrac{\delta}{\delta {#1}}}
\newcommand {\deedee}[1]{\frac{\partial}{\partial {#1}}}
\newcommand {\ldeedee}[1]{\lfrac{\partial}{\partial {#1}}}
\newcommand {\ddl}{\frac{d}{d\lambda}}
\newcommand {\ddz}{\frac{d}{dz}}
\newcommand {\lddz}{\lfrac{d}{dz}}
\newcommand {\deltil}[1]{\tilde{\partial}_{#1}}
\newcommand \zb{\bar{z}}
\newcommand \delz{\del{z}}
\newcommand \delzb{\del{\bar{z}}}
\newcommand \hnab{{\hat \nabla}}

\newcommand \bra[1]{\left< {#1} \,\right\vert}
\newcommand \ket[1]{\left\vert\, {#1} \, \right>}
\newcommand \braket[2]{\hbox{$\left< {#1} \,\vrule\, {#2} \right>$}}

\newcommand \bA{{\bf A}}
\newcommand \bB{{\bf B}}
\newcommand \bC{{\bf C}}
\newcommand \bL{{\bf L}}
\newcommand \bM{{\bf M}}
\newcommand \bN{{\bf N}}
\newcommand \bR{{\bf R}}
\newcommand \bZ{{\bf Z}}
\newcommand \bW[2]{{\bf W}^{(#1)}_{#2}}
\newcommand \bWz[1]{{\bf W}^{(#1)}(\la\pr)}
\newcommand \bWw[1]{{\bf W}^{(#1)}(\la)}
\newcommand \bWa[1]{{\bf W}^{(#1)}}
\newcommand \cG{{\cal G}}
\newcommand \cH{{\cal H}}
\newcommand \cJ{{\cal J}}
\newcommand \Jt{\tilde{J}}
\newcommand \cJt{\tilde{\cal J}}
\newcommand \cL{{\cal L}}
\newcommand \cO{{\cal O}}
\newcommand \cR{r}
\newcommand \cS{{\cal S}}
\newcommand \RO{{\cal O}}
\newcommand \RD{{\cal D}}
\newcommand \Sk[1]{S^{(#1)}}
\newcommand \Wz[1]{W^{(#1)}(z)}
\newcommand \Wtz[1]{\tilde{W}^{(#1)}(z)}
\newcommand \W[2]{W^{(#1)}_{#2}}
\newcommand \w[2]{w^{(#1)}_{#2}}
\newcommand \Wba[1]{\bar{W}^{(#1)}}
\newcommand \Wt[2]{\tilde{W}^{(#1)}_{#2}}
\newcommand \cW[2]{{\cal W}^{(#1)}_{#2}}
\newcommand \cw[2]{w^{(#1)}_{#2}}
\newcommand \cWb[2]{\bar{\cal W}^{(#1)}_{#2}}
\newcommand \xiir[2]{\xi^{({#1})}_{#2}}
\newcommand \etan[2]{\eta^{({#1})}_{#2}(x)}
\newcommand \etanx[1]{\eta^{({#1})}(x)}
\newcommand \xiz[1]{\xi^{({#1})}(z)}
\newcommand \psidag[1]{{\psi^{{({#1})}}}^{\dagger}}
\newcommand \Pb{\bar{P}}
\newcommand \Qb{\bar{Q}}
\newcommand \Wb{\bar{W}}
\newcommand \wb{\bar{w}}
\newcommand \xb{\bar{x}}
\newcommand \ab{\bar{a}}
\newcommand \tb{\bar{t}}
\newcommand \hg{{\hat g}}

\newcommand {\tr}{\mbox{tr}\,}
\newcommand \Tr {\mbox {Tr}}
\newcommand \any {\mbox {~any~}}
\newcommand \with {\mbox {~with~}}
\newcommand \for {\mbox {~for~}}
\newcommand \where {\mbox {~where~}}
\newcommand \here {\mbox {~here~}}
\newcommand \ando {\mbox {~and~}}
\newcommand \oa {\mbox {~or~}}
\newcommand \diag {\mbox{diag}}
\newcommand \etc {\mbox{~etc~}}
\newcommand \mod {\mbox{\,mod~}}
\newcommand \const {\mbox{const.\,}}

\newcommand \la{\lambda}
\newcommand \sig{\sigma}
\newcommand \sigpr{\sigma^{\prime}}
\newcommand \xpr{x^{\prime}}
\newcommand \ypr{y^{\prime}}
\newcommand \eps{\epsilon}
\newcommand \han{\frac{1}{2}}
\newcommand \san{\frac{1}{3}}
\newcommand \reg{\mbox{reg.}}

\newcommand \cd{\cdots}
\newcommand \dd{\ddots}
\newcommand \vd{\vdots}
\newcommand \pr{^\prime}
\newcommand \prr{^{\prime\prime}}
\newcommand \ol[1]{\overline{#1}}

\newcommand \uaa{\hspace{-7.2mm}}

\newcommand \mti[1]{\mbox{\tiny {#1}}}
\newcommand \msc[1]{\mbox{\scriptsize {#1}}}
\newcommand \mfo[1]{\mbox{\footnotesize {#1}}}
\newcommand \msm[1]{\mbox{\small {#1}}}
\newcommand \mno[1]{\mbox{\normalsize {#1}}}
\newcommand \mla[1]{\mbox{\large {#1}}}
\newcommand \mLa[1]{\mbox{\Large {#1}}}
\newcommand \mLA[1]{\mbox{\LARGE {#1}}}
\newcommand \mhu[1]{\mbox{\huge {#1}}}
\newcommand \mHu[1]{\mbox{\Huge {#1}}}

\newcommand {\lno}
  {
    \begin{array}{r}
    \raisebox{-1.2mm}{\msc{$\circ$}} \\ \raisebox{3.2mm}{\msc{$\circ$}}
    \end{array}
   \!\!}
\newcommand {\rno}
  {\!\!
    \begin{array}{l}
    \raisebox{-1.2mm}{\msc{$\circ$}} \\ \raisebox{3.2mm}{\msc{$\circ$}}
    \end{array}
   }
\newcommand {\lp}{\left(}
\newcommand {\rp}{\right)}

\newcommand {\ud}{\updownarrow}


Great progress in the formulation of 2d quantum gravity has been made since
the equivalence of 2d quantum gravity and
dynamical triangulation (DT)\cite{kkm} was established. \cite{matrix}
\cite{d} \cite{fkn} \cite{dvv} \cite{ap} \cite{yoneya} Conformally invariant
matters with Virasoro central charge $c_M \leq 1$ can be consistently
coupled to 2d gravity \cite{dkd} \cite{kpz}
and correlation functions computed in both continuum
and DT approaches coincide.\cite{gl} The partition function for fixed
area $A$ is given by \cite{dkd}
\beq
Z(A) = \mbox{const.} \, A^{\gamma_{str}(c_M,h)-3} \, e^{-\frac{\mu}{2\pi}A},
\label{Z1.1}
\eeq
where $\gamma_{str}$ is the string susceptibility \cite{dkd} \cite{kpz}
\beq
\gamma_{str}(c_M,h) = \lfrac{1}{12} [ c_M-25-\sqrt{(25-c_M)(1-c_M)}](1-h)+2,
\label{G1.2}
\eeq
$h$ is the number of handles of the closed orientable surface and $\mu$ is
the renormalized cosmological constant.  Note that $\gamma_{str}$ is real
for $c_M \leq 1$.

If $c_M > 1$, however, $\gamma_{str}$ becomes imaginary and the continuum
approach fails in quantization of 2d gravity.  The system is in a branched
polymer phase and a full-fledged surface theory has yet to be  found.
However, there have been several attempts to circumvent the $c_M =1$ barrier,
{\it e.g.}, the extrinsic curvature term, W gravity, quantum groups,
{\it etc.}
In this letter we will consider the $R^2$ gravity theory.  As is well-known,
surfaces have many spikes corresponding to large local fluctuations.
At the ends of spikes large scalar curvature $R$ is localized.  To smooth
out surfaces we have to disperse localized curvature all over the surface.
Adding the $R^2$ term to the action will suppress wild fluctuations of
$R$.\footnote{This term was first introduced in \cite{bkkm}.}
We will investigate the $R^2$ gravity within the framework of conformally
invariant field theory and show that the $R^2$ term affects the short
distance property of surfaces and the problem of the branched polymer
is avoided.

We will consider a gravity theory described by an action
\begin{equation}
\lfrac{1}{8\pi} \int d^2 x ( \frac{1}{4m^2} R^2 + 4\mu_0 ),
\label{S2.1}
\end{equation}
where $R$ is the scalar curvature, $\mu_0$ the bare cosmological constant
and $m$ a coupling constant of mass dimension 1.  This is a fourth-order
derivative theory and rather difficult to work with.  We will instead
introduce an auxiliarly field $\chi$ and define our theory by an action:
\footnote{Gaussian integration over $\chi$ in the conformal gauge (\ref{G2.5})
will formally yield a factor $\prod_x exp \{-\han \phi(x) \}$, which is
equal to $exp \{ \frac{-1}{48\pi} S_L(\phi;\hg) \} $ when evaluated
by using Heat kernel method.
This cancels RHS of eq (\ref{A2.12}) hence the theories (\ref{S2.1})
and (\ref{S2.2}) are equivalent. }
\begin{equation}
S_{\chi}(\chi;g) = \lfrac{1}{8\pi} \int d^2 x \sqrt{g}
                       (-i R \chi + m^2 \chi^2 + 4\mu_0 ).
\label{S2.2}
\end{equation}
The partition function is given by
\begin{equation}
Z=\int \lfrac{\RD g \RD \chi \RD X}{vol(Diff)}
             e^{-S_{\chi}(\chi;g)-S_{M}(X^i;g)},
\label{Z2.3}
\end{equation}
where $S_M(X^i;g)$ is the matter action given by
\beq
S_{M}(X^i;g) = \lfrac{1}{8\pi} \int d^2 x \sqrt{g} g^{\mu \nu} \del{\mu}X^i
               \del{\nu}X^i .
\eeq
Here $X^i (i=1,2, \cdots ,d)$ is a scalar field taking values in $\bR^d$.
The action and the integration measure in the partition function
(\ref{Z2.3}) are invariant under diffeomorphisms.  Hence the measure is
divided by the volume of the diffeomorphisms, which is denoted by $vol(Diff)$.

In the conformal gauge we set
\beq
g_{\mu \nu} = \hg_{\mu \nu} (x;\tau) e^{\phi (x)},
\label{G2.5}
\eeq
where $\phi(x)$ is a conformal mode and $\hg_{\mu \nu}(x;\tau)$ is a
background metric which depends on additional parameters, moduli $\tau$.
The integration measure $\RD_g X$ is induced from the following metric in the
functional space of $X^i$
\beq
\| \delta X^i \|_g ^2 = \int d^2 x \sqrt{g} (\delta X^i)^2
                      = \int d^2 x \sqrt{\hg} e^{\phi} (\delta X^i)^2 .
\label{M2.6}
\eeq
Although $S_M (X^i;g) = S_M (X^i;\hg) $, {\it i.e.} $S_M$ is independent of
$\phi (x)$, $\RD _g X$ depends on $\phi (x)$ via conformal anomaly
\beq
\RD_g X = \RD_{\hg} X \ exp \{ \lfrac{c_M}{48\pi} S_L (\phi;\hg) \}.
\eeq
Here $c_M = d$ and $S_L(\phi;\hg)$ is the Liouville action:
\beq
S_L (\phi;\hg) = \int d^2 x \sqrt{\hg} \ (\han \hg^{\mu \nu} \del{\mu}\phi
                 \del{\nu}\phi + {\hat R}\phi + 2\mu_1 e^{\phi}).
\eeq
The measure $\RD_g g$ is defined by the metric
\beq
\|\delta g \|_g ^2 = \int d^2 x \sqrt{g} \ (g^{\mu \nu} g^{\lambda \rho}
                  + a_0 \, g^{\mu \lambda} g^{\nu \rho}) \delta g_{\mu \lambda}
                        \delta g_{\nu \rho},
\eeq
where $a_0 $ is a constant.  This induces the following metric in the
functional
space of $\phi$
\beq
\|\delta \phi \|_g ^2 = 2(1+2a_0) \int d^2 x \sqrt{\hg} e^{\phi}(\delta
                       \phi)^2 \propto \int d^2 x \sqrt{g} (\delta
                       \phi)^2.
\eeq
This metric is identical with the one in eq(\ref{M2.6}), and we have
\beq
\RD_g \phi = \RD_{\hg} \phi \ exp \{ \lfrac{1}{48\pi} S_L(\phi;\hg) \}.
\eeq
Similarly we obtain
\beqa
\RD_g \chi & = & \RD_{\hg} \chi \ exp \{ \lfrac{1}{48\pi}S_L(\phi;\hg) \},
\label{A2.12} \\
\RD_g b \RD_g c & = & \RD_{\hg} b \RD_{\hg} c
                \      exp \{ \lfrac{-26}{48\pi} S_L(\phi;\hg) \},
\eeqa
where $b$ ($=b_{\mu \nu}$) and $c$ ($=c^{\mu}$) are the reparametrization
ghosts.

The partition function (\ref{Z2.3}) now takes the form
\beq
Z=\int d \tau \int \RD_\hg \phi \RD_\hg \chi \RD_\hg X \RD_\hg b \RD_\hg c
 \      exp \{ -S_0(\phi,\chi;\hg)-S_M(X^i;\hg)-S_{gh}(b,c;\hg) \},
\label{Z2.14}
\eeq
where
\beq
S_0(\phi,\chi;\hg) = S_{\chi} (\chi;e^{\phi}\hg) - \lfrac{d-24}{48\pi}
                   S_L(\phi;\hg),
\eeq
and $S_{gh}(b,c;\hg)$ is the action for the ghosts.
By using
\beq
\sqrt{g} R = \sqrt{\hg}({\hat R} - \hnab ^{\mu} \hnab_{\mu} \phi),
\label{F2.16}
\eeq
we have
\beqa
S_0(\phi,\chi;\hg) &  =  & \lfrac{1}{8\pi} \int d^2 x \sqrt{\hg} \{ \hg^{\mu
\nu}(-i \del{\mu}\phi
  \del{\nu}\chi + \lfrac{24-d}{12} \del{\mu}\phi \del{\nu}\phi ) \n
 & & \quad \quad -i {\hat R}\chi + \lfrac{24-d}{6} {\hat R} \phi
            + m^2 e^{\phi} \chi^2  + 4\mu e^{\phi} \}.
\label{S2.17}
\eeqa
The last two terms should be renormalized.  To fix the forms of renormalized
operators we note that the partition function (\ref{Z2.14}) should be
invariant under the transformation given by \cite{dkd}
\beqa
\hg_{\mu \nu}(x) & \rightarrow & \hg_{\mu \nu}(x) e^{\sigma (x)}, \n
\phi (x)         & \rightarrow & \phi (x) - \sigma (x),
\label{W2.18}
\eeqa
because the original theory (\ref{Z2.3}) is defined in terms of $\phi$ and
$\hg_{\mu \nu}$ through the product $e^{\phi (x)} \hg_{\mu \nu}(x)$.
It is easy to check that if $m^2 = \mu = 0$, eq(\ref{Z2.14}) is invariant
under the transformation (\ref{W2.18}).  Therefore the terms $\sqrt{\hg}
e^{\phi} \chi^2$ and
$\sqrt{\hg}e^{\phi}$ should be renormalized so that they are invariant
separately.   Equivalently
we may regard these terms as perturbations and demand that these operators
have conformal weight (1,1).   The unperturbed $(m^2 = \mu = 0)$  theory is
conformally invariant and the stress tensor $T_{zz}$ is given by
\beq
T_{zz}= \lfrac{i}{2} : \del{z} \chi \del{z} \phi : - \lfrac{24-d}{24}
       : (\del{z}\phi)^2: - \lfrac{i}{2} \del{z}^2 \chi
       + \lfrac{24-d}{12} \del{z}^2 \phi.
\label{T2.19}
\eeq
The operator product expansion (OPE) of $T_{zz}$ can be computed by using
\beqa
\chi(z) \chi(w) & = & - \lfrac{24-d}{3} \ln (z-w), \n
\chi(z) \phi(w) & = & -2i \ln (z-w), \n
\phi(z) \phi(w) & = & 0,
\label{OPE2.20}
\eeqa
and we find that the central charge of Virasoro algebra generated by
$T_{zz}$ is given by $26-d$, which indeed cancels the central charges
of the matter and ghosts.

Instead of renormalizing $\sqrt{\hg} e^\phi $ and $\sqrt{\hg} e^\phi \chi^2 $
separately, we will consider a general operator
\beq
V_n ^{(0)} (\phi,\chi) = \sqrt{\hg} e^\phi \chi^n .
\eeq
To renormalize $V_n ^{(0)}$ we consider the following vertex operator
\beq
V(\phi,\chi;\alpha,\beta) = : e^{\alpha \phi + i \beta \chi } : ,
\eeq
whose conformal weight is found, by using eqs (\ref{T2.19}) and
(\ref{OPE2.20}), to be
\beq
\Delta_0 (\alpha,\beta) = \alpha+2\alpha\beta+\lfrac{24-d}{6}\beta^2.
\eeq
Demanding $\Delta_0 (\alpha,\beta)=1$ yields
\beq
\alpha = \alpha (\beta) =  \sfrac{1-\frac{24-d}{6}\beta^2}{1+2\beta}
        =  1-2\beta+\lfrac{d}{24} \sum_{n=2}^{\infty}(-2 \beta)^n ,
\eeq
and by expanding $V(\phi,\chi;\alpha (\beta),\beta)$ in powers of $\beta$
we obtain an infinite set of (1,1) operators $V_n(\phi,\chi)$,
\beqa
V(\phi,\chi;\alpha(\beta),\beta) & = &
        \sqrt{\hg} : exp \{ \phi+\beta (i\chi-2\phi)+ \lfrac{d}{24}
        \sum_{n=2}^{\infty}(-2\beta)^n \phi \} : , \n
& = & \sum_{n=0}^{\infty} \lfrac{(i\beta)^n}{n!} V_n(\phi,\chi).
\eeqa
Renormalization modifies $V_n ^{(0)} (\phi,\chi)$ into $V_n (\phi,\chi)$,
the first few of which read
\beqa
V_0(\phi,\chi) & = & \sqrt{\hg} : e^{\phi} :, \n
V_1(\phi,\chi) & = & \sqrt{\hg} : e^{\phi} (\chi+2i \phi) :, \n
V_2(\phi,\chi) & = & \sqrt{\hg} : e^{\phi} \{
                          (\chi+2i \phi)^2-\lfrac{d}{3}\phi \} :, \n
V_3(\phi,\chi) & = & \sqrt{\hg} : e^{\phi} \{
               (\chi+2i \phi)^3 -d (\chi+2i \phi) \phi -2id \, \phi \} :.
\eeqa
The cosmological term is not renormalized because $\phi (z) \phi(w)=0$.
By taking into account renormalization of $\sqrt{\hg} e^\phi \chi^2$ and
$\sqrt{\hg} e^\phi$ and defining a new field $F$ according to
\beq
F = \chi + 2i \phi,
\eeq
we rewrite the action (\ref{S2.17}) in the form
\beqa
S(\phi,F;\hg) & = & S_0(\phi,F-2i \phi;\hg)   \n
              & = & \lfrac{1}{8\pi} \int d^2 x \sqrt{\hg}
                          \{ \hg^{\mu \nu} (-i \del{\mu} \phi \del{\nu} F
                       -\lfrac{d}{12} \del{\mu} \phi \del{\nu} \phi ) \n
              &   & -i {\hat R} F + \lfrac{12-d}{6} {\hat R} \phi
                    + m^2 :e^\phi (F^2 -\lfrac{d}{3} \phi): + 4\mu e^\phi \}.
\label{S2.28}
\eeqa

The partition function for fixed area $A$ is given by
\beqa
Z(A;m,\mu) & = & \int d\tau \int \RD_{\hg} \phi \RD_{\hg} F \RD_{\hg} X
                 \RD_{\hg} b \RD_{\hg} c \
                   \delta ( \int d^2x \sqrt{\hg} e^{\phi}-A) \n
           & &      exp \{ -S(\phi,F;\hg)-S_M(X^i;\hg)-S_{gh}(b,c;\hg) \}.
\label{Z2.29}
\eeqa
We integrate over the zero mode of $\phi$.  We define
\beq
\phi (x) = \tilde{\phi}(x) + \phi_0 ,
\eeq
where $\phi_0$ is a constant.
By using
\beq
\int d^2 x \sqrt{\hg} {\hat R} = 8\pi (1-h)
\eeq
($h$ is the number of handles) and
\beq
\delta(\int d^2x \sqrt{\hg} e^\phi - A) =
 \lfrac{1}{A} \delta (\phi_0+\ln \{ \lfrac{1}{A} \int d^2x \sqrt{\hg}
                              e^{{\tilde \phi}} \}),
\eeq
we integrate over $\phi_0$ to obtain
\beq
Z(A;m,\mu)= A^{-1+\frac{d-12}{6}(1-h)} \, A^{\frac{d}{24\pi}m^2A}
            \,  e^{-\frac{1}{2\pi}\mu A} \, Y(m^2A).
\label{Z2.33}
\eeq
Here $Y(m^2A)$ is defined by the following integral
\beqa
Y(m^2A) & = & \int d\tau \int \RD_{\hg} \tilde{\phi} \RD_{\hg} F \RD_{\hg} X
                   \RD_{\hg} b \RD_{\hg} c \
            (G[\tilde{\phi}])^{\frac{12-d}{6}(1-h)-\frac{d}{24\pi}m^2A} \n
& & exp [ -\lfrac{1}{8\pi} \int d^2x \sqrt{\hg} \
                  \{\hg^{\mu \nu} (-i \del{\mu} \tilde{\phi} \del{\nu} F
                 -\lfrac{d}{12} \del{\mu} \tilde{\phi} \del{\nu} \tilde{\phi})
                -i \hat{R}F \n
& & \quad +\lfrac{12-d}{6} \hat{R} \tilde{\phi} \}
      -\lfrac{1}{8\pi} m^2A(G[\tilde{\phi}])^{-1} \int d^2x \sqrt{\hg}
          : e^{\tilde{\phi}}  \{F^2-\lfrac{d}{3} \tilde{\phi} \}: \n
& & \quad -S_M(X^i;\hg)-S_{gh}(b,c;\hg) ],
\label{Y2.35}
\eeqa
where
\beq
G[\tilde{\phi}] = \int d^2x \sqrt{\hg} e^{\tilde{\phi}}.
\eeq
We wil show below that
\beq
Y(m^2A)  \sim  \mbox{const.} exp \ \{ -\lfrac{2\pi}{m^2A} (1-h)^2 \}
         \quad \mbox{for} \ \  m^2A \rightarrow 0.  \label{Y2.38}
\eeq
 From eqs (\ref{Z2.33}) and (\ref{Y2.38}) we obtain
the following asymptotic form for $Z(A;m,\mu)$
\beqa
Z(A;m,\mu) & \sim & \mbox{const.} A^{\gamma_{str}(h,m^2A)-3}
          \, exp \{ -\lfrac{2\pi}{m^2A}(1-h)^2 \} \,
     e^{-\frac{1}{2\pi} (\mu + b_0 m^2) A} \n
           & & \quad \quad   \mbox{for} \ \ m^2A \rightarrow 0,
\label{Z2.40}
\eeqa
where
$b_0$ is a constant and
\beq
\gamma_{str}(h,m^2A)= 2+\lfrac{d-12}{6}(1-h)+\lfrac{d}{24\pi}m^2A
\label{G2.41}
\eeq
is the string susceptibility.  Note that this depends on the area $A$ and
$\gamma_{str}$ is real for any value of $d$.

To estimate the limit $m^2A \rightarrow 0$, we have to pay special
attention to the $F$ zero mode, $F_0$.  If we formally set $m^2A=0$ in eq
(\ref{Y2.35}), the potential for $F_0$ is given by
\beq
-\lfrac{i}{8\pi} \int d^2x \sqrt{\hg} \hat{R} F_0 = -i (1-h) F_0 .
\eeq
This is not bounded even if $F_0$ is analytically continued to $iF_0$
and the integral is not well-defined.
In order to take the correct $m^2A \rightarrow 0$ limit, we set
\beq
F(x) = \tilde{F}(x) + \lfrac{4\pi i (1-h)}{m^2A},
\eeq
and substitute this into eq(\ref{Y2.35}).
Then $F$ dependent part of eq (\ref{Y2.35}) can be rewritten as follows
\beqa
-\lfrac{1}{8\pi} \int d^2x \sqrt{\hg} (-i \hg^{\mu \nu} \del{\mu} \tilde{\phi}
             \del{\nu}F-i \hat{R} F ) & - & \lfrac{1}{8\pi}
             \lfrac{m^2A}{G[\tilde{\phi}]} \int d^2x \sqrt{\hg}
           : e^{\tilde{\phi}} F^2: \n
= -\lfrac{1}{8\pi} \int d^2x \sqrt{\hg} (-i \hg^{\mu \nu} \del{\mu}
         \tilde{\phi} \del{\nu} \tilde{F} & - & i: \tilde{F} \{ \hat{R} -
          \lfrac{8\pi(1-h)e^{\tilde{\phi}}}{G[\tilde{\phi}]} \}: ) \n
- \lfrac{1}{8\pi} \lfrac{m^2A}{G[\tilde{\phi}]} \int d^2x \sqrt{\hg}
           : e^{\tilde{\phi}} \tilde{F}^2: & - & \lfrac{2\pi (1-h)^2}{m^2A}.
\eeqa
The first line on the RHS is invariant under the transformation
\beq
\tilde{F}(x) \rightarrow \tilde{F}(x) + \mbox{const},
\eeq
and so the volume of $F_0$ integration factors out in the limit
$m^2A \rightarrow 0$.

This result can also be derived as follows.
By formally integrating over $F$ in eq (\ref{S2.28}) we obtain an effective
action (see footnote~2),
\beqa
S_{eff} & = & \lfrac{1}{8\pi} \int d^2x \sqrt{\hg} [ \lfrac{1-d}{12}
      \hg^{\mu \nu}\del{\mu}\phi \del{\nu}\phi +\lfrac{13-d}{6}\hat{R}\phi -
      \lfrac{d}{3}m^2\phi e^\phi +4\mu' e^\phi ] \n
&  & + \lfrac{1}{8\pi} \int d^2x \lfrac{1}{4m^2} \sqrt{g} R^2 .
\eeqa
In the limit $m \rightarrow 0$ we end up with a constraint $R=0$,
which is inconsistent with $\int d^2x \sqrt{g}R=8\pi (1-h)$.
Indeed a correct constraint has to be
\beq
R= \rho= \lfrac{8\pi}{A}(1-h).
\label{R.47}
\eeq
Hence we obtain
\footnote{When $m \rightarrow 0$, $(32\pi m^2)^{-1} \int d^2x \sqrt{g}
(R-\rho)^2$ is equivalent to $\int d^2x \sqrt{g} \Phi (R-\rho)$,
where $\Phi$ is a Lagrange multiplier field.  The latter action was studied in
\cite{cham}.  Since $\rho = 8\pi (1-h) /\int d^2x \sqrt{g}$ (eq(\ref{R.47})),
this action is non-local in contrast with $(32\pi m^2)^{-1} \int d^2x \sqrt{g}
R^2$.}
\beqa
\lfrac{1}{32\pi m^2}\int d^2x \sqrt{g}R^2 & = &
             \lfrac{1}{32\pi m^2}\int d^2x \sqrt{g} [(R-\rho)^2+2\rho R-
              \rho^2 ] ,  \n
& = & \lfrac{1}{32\pi m^2} [\int d^2x \sqrt{g}(R-\rho)^2+16\pi \rho (1-h)-
            \rho^2A ] ,   \n
& = & \lfrac{1}{32\pi m^2}\int d^2x \sqrt{g}(R-\rho)^2 +
            \lfrac{2\pi (1-h)^2}{m^2A}.
\eeqa

As for the $m^2A \rightarrow \infty$ limit, we cannot directly
investigate eq(\ref{Y2.35}) because the $\chi$ mass term is a perturbation.
Actually the potential in eq (\ref{S2.28}) (after analytical
continuation $F \rightarrow iF$)
\beq
\hat{R}F+\lfrac{12-d}{6}\hat{R}\phi
       +m^2 : e^\phi (-F^2-\lfrac{d}{3}\phi):+ 4\mu e^\phi
\eeq
is not bounded below for $F \rightarrow \pm \infty$ or
in the IR limit $\phi \rightarrow +\infty$ if $d > 0$.
By a physical consideration, however, we expect that the scalar
field $\chi$ with infinite mass will decouple from the rest of the system.
In other words the $R^2$ term will not affect the long distance physics
and the asymptotic behavior of $Z(A;m,\mu)$ will be given by
\beq
Z(A;m,\mu) \sim \mbox{const.} \, A^{\gamma_{str}^\infty} e^{-\frac{\mu}{2\pi}
               A} \quad \quad \mbox{for} \quad m^2A \rightarrow \infty,
\label{Z2.47}
\eeq
where $\gamma_{str}^\infty$ is the string susceptibility of 2d gravity
without the $R^2$ term\cite{dkd} \cite{kpz}:
\beq
\gamma_{str}^\infty = \lfrac{1}{12} [d-25-\sqrt{(25-d)(1-d)}](1-h)+2.
\eeq

 From the above results (\ref{Z2.40}) and (\ref{Z2.47}), we can draw the
following
conclusion.  Due to the exponential factor in eq (\ref{Z2.40}) the
partition function for $A \ll 1/m^2$ is highly suppressed.  Therefore
at a length scale smaller than $1/m$ surfaces will be smooth.  On the
contrary at a length scale much larger than $1/m$ surfaces will have spiky
structures and if $d > 1 $, they will look like branched polymers.

We can study roughness of surfaces by marking a point P on a surface $\Sigma$
and considering a connected region consisting of points within geodesic
distance $r$ from P.  If $\Sigma$ is flat, this region will be a disk.  If
$\Sigma$ is fluctuating and has spikes, this region will have many
boundaries.  Recently the number of boundaries whose length is in between $l$
and $l+dl$, $f(l,r)dl$, was derived in pure gravity without the $R^2$
term.\cite{kkmw}
\beq
f(l,r) dl \sim \lfrac{1}{r^2} \, e^{-l/r^2} \,
      [\left( \lfrac{l}{r^2} \right) ^{-\frac{5}{2}}+ \han
      \left( \lfrac{l}{r^2} \right) ^{-\frac{3}{2}}+\lfrac{14}{3} \left(
      \lfrac{l}{r^2} \right) ^\han] dl
\eeq
This is singular for $l \rightarrow 0$ and so there are an infinite number
of very thin spikes.  We find that
the number of boundaries $\int_0^\infty dl f(l,r)$ and the total length
of boundaries $\int_0^\infty dl f(l,r)l$ are divergent.  If we introduce
the $R^2$ term to the action, spikes of size $l$ smaller than $1/m$ may be
suppressed in the same fashion as surfaces of area A smaller
than $1/m^2$ are.  This deserves a further study.

Finally let us give a few remarks.
\begin{enumerate}
\item  $R^2$ gravity theory does not satisfy positivity.  This can be easily
seen by analytic continuation $\chi \rightarrow i \chi$.  We have a particle
of mass $\sim m$ with a wrong-sign kinetic term.
\item If we consider a $\chi$ potential $V(\chi)$ different from $m^2\chi^2$,
we obtain a model in a different universality class.  For an action
\beq
S = \lfrac{1}{8\pi} \int d^2x \sqrt{g}(R\chi-m^2\chi^{2p}+4\mu_0),
\eeq
the partition function behaves  as
\beqa
Z(A) & \sim & A^{\gamma_{str}'-3} \, exp \{ -\lfrac{2p-1}{2p}(1-h)
       \{ \lfrac{4\pi (1-h)}{pm^2A} \}^{1/(2p-1)} \} \, e^{-\frac{\mu}{2\pi}A}
  \n
 & & \mbox{for} \quad m^2A \rightarrow 0.
\eeqa
\item
We can slightly generalize the model(\ref{S2.2}) by introducing a kinetic term
for $\chi$
\beq
S_{\chi}(\chi;g,\xi)= \lfrac{1}{8\pi} \int d^2x \sqrt{g} ( \lfrac{1}{4\xi^2}
                  g^{\mu \nu} \del{\mu} \chi \del{\nu} \chi + m^2 \chi^2
                  -i R \chi + 4 \mu_0 ),
\label{S3.1}
\eeq
where $\xi$ is a parameter and in the limit $\xi \rightarrow \infty$ this
action coincides with eq (\ref{S2.2}).\footnote{  This model has a geometrical
meaning.  In the previous argument the world-sheet geometry was assumed to be
torsionless.  The model (\ref{S3.1}) can be shown to be identical with the
$R^2$ gravity theory with torsion\cite{kuma} with the spin connection
$\omega_\mu$ integrated out.}

Quantization of the model(\ref{S3.1}) proceeds in a parallel fashion
with the previous
analysis of eq (\ref{S2.2}).  The main difference is that the cosmological
term is renormalized $\sqrt{\hg} : exp (\alpha \phi):$ and there are two
possible values of $\alpha = \alpha_\pm $, where
\beq
\alpha_{\pm} =
       \lfrac{1}{12} [24-d+12\xi^2 \pm \sqrt{(12\xi^2-d)(24+12\xi^2-d)}] .
\eeq
The values of $\alpha_{\pm}$ are real and positive if
\beq
d \leq 12\xi^2 \quad \quad \mbox{or} \quad \quad d \geq 12\xi^2 + 24 .
\eeq
We will restrict attention to the region $d \leq 12\xi^2$ which is accessible
from the semiclassical limit.  In the limit $\xi \rightarrow \infty$ or
$d \rightarrow -\infty$,
$\alpha_- $ approaches 1 but $\alpha_+$ blows up.
Therefore we should choose $\alpha_-$ as the correct value.
The renormalized $\chi$ mass term is given by
\beqa
m^2 \sqrt{\hg} : e^{\alpha_- \phi}  & [ & \{ \chi -2i \xi^2 \left( 1 -
        \sqrt{\lfrac{24-d+12\xi^2}{-d+12\xi^2}} \right) \phi \} ^2    \n
 & - & \lfrac{4\xi^2 d}{-d+12\xi^2} \sqrt{\lfrac{24-d+12\xi^2}{-d+12\xi^2}}
                    \phi \, ]: .
\label{M3.48}
\eeqa
By an analysis of the partition function for fixed area,
$Z(A;\xi,m,\mu)$,  we obtain an asymptotic form for $m^2A \rightarrow
\infty$ which is the same as eq(\ref{Z2.40})
with the string susceptibility $\gamma_{str}$ (\ref{G2.41})
modified according to
\beqa
\gamma_{str}(d,h,\xi,m^2A) & = & 2 - \lfrac{2}{\alpha_-} \left(
              \lfrac{24-d+12\xi^2}{12} - \xi^2
              \sqrt{\lfrac{24-d+12\xi^2}{-d+12\xi^2}} \right) (1-h) \n
& + & \lfrac{\xi^2 d}{2\pi \alpha_- (-d+12\xi^2)}
         \sqrt{\lfrac{24-d+12\xi^2}{-d+12\xi^2}} m^2A .
\eeqa
\end{enumerate}

\vspace{1cm}
We are grateful to Y.~Okamoto for discussions and to N.~Kawamoto for
comments.

\end{document}